\documentclass[aps,prd,twocolumn,superscriptaddress,noshowkeys,noshowpacs]{revtex4-1}

\usepackage{amsmath,amssymb,color,graphicx,bm,braket,nicefrac}
\usepackage[toc,page]{appendix}

\newcommand{\bern}{\affiliation{Albert Einstein Center for Fundamental Physics, University of Bern, Bern, Switzerland}}
\newcommand{\bologna}{\affiliation{Department of Physics and Astrophysics, University of Bologna and INFN-Bologna, Bologna, Italy}}
\newcommand{\chicago}{\affiliation{Department of Physics \& Kavli Institute of Cosmological Physics, University of Chicago, Chicago, IL, USA}}
\newcommand{\coimbra}{\affiliation{Department of Physics, University of Coimbra, Coimbra, Portugal}}
\newcommand{\columbia}{\affiliation{Physics Department, Columbia University, New York, NY, USA}}
\newcommand{\lngs}{\affiliation{INFN-Laboratori Nazionali del Gran Sasso and Gran Sasso Science Institute, L'Aquila, Italy}}
\newcommand{\mainz}{\affiliation{Institut f\"ur Physik \& Exzellenzcluster PRISMA, Johannes Gutenberg-Universit\"at Mainz, Mainz, Germany}}
\newcommand{\heidelberg}{\affiliation{Max-Planck-Institut f\"ur Kernphysik, Heidelberg, Germany}}
\newcommand{\munster}{\affiliation{Institut f\"ur Kernphysik, Wilhelms-Universit\"at M\"unster, M\"unster, Germany}}
\newcommand{\nikhef}{\affiliation{Nikhef and the University of Amsterdam, Science Park, Amsterdam, Netherlands}}
\newcommand{\nyuad}{\affiliation{New York University Abu Dhabi, Abu Dhabi, United Arab Emirates}}
\newcommand{\purdue}{\affiliation{Department of Physics and Astronomy, Purdue University, West Lafayette, IN, USA}}
\newcommand{\rpi}{\affiliation{Department of Physics, Applied Physics and Astronomy, Rensselaer Polytechnic Institute, Troy, NY, USA}}
\newcommand{\rice}{\affiliation{Department of Physics and Astronomy, Rice University, Houston, TX, USA}}
\newcommand{\stockholm}{\affiliation{Oskar Klein Centre, Department of Physics, Stockholm University, AlbaNova, Stockholm, Sweden}}
\newcommand{\subatech}{\affiliation{SUBATECH, Ecole des Mines de Nantes, CNRS/In2p3, Universit\'e de Nantes, Nantes, France}}
\newcommand{\torino}{\affiliation{INFN-Torino and Osservatorio Astrofisico di Torino, Torino, Italy}}
\newcommand{\ucla}{\affiliation{Physics \& Astronomy Department, University of California, Los Angeles, CA, USA}}
\newcommand{\ucsd}{\affiliation{Department of Physics, University of California, San Diego, CA, USA}}
\newcommand{\wis}{\affiliation{Department of Particle Physics and Astrophysics, Weizmann Institute of Science, Rehovot, Israel}}
\newcommand{\zurich}{\affiliation{Physik-Institut, University of Zurich, Zurich, Switzerland}}

\newcommand{\be}{\begin{equation}}
\newcommand{\ee}{\end{equation}}
\newcommand{\1}[1]{\, \mathrm{#1}} 

\begin{document}
\title{Results from a Calibration of XENON100 Using a Source of Dissolved Radon-220}
\author{E.~Aprile}\columbia
\author{J.~Aalbers}\nikhef
\author{F.~Agostini}\lngs\bologna
\author{M.~Alfonsi}\mainz
\author{F.~D.~Amaro}\coimbra
\author{M.~Anthony}\columbia
\author{F.~Arneodo}\nyuad
\author{P.~Barrow}\zurich
\author{L.~Baudis}\zurich
\author{B.~Bauermeister}\stockholm\mainz
% ...
\collaboration{XENON Collaboration}\email{xenon@lngs.infn.it}\noaffiliation

\begin{abstract}
A $^{220}$Rn source is deployed on the XENON100 dark matter detector in order to address the challenges in calibration of tonne-scale liquid noble element detectors. We show that the $^{212}$Pb beta emission can be used for low-energy electronic recoil calibration in searches for dark matter. The isotope spreads throughout the entire active region of the detector, and its activity naturally decays below background level within a week after the source is closed. We find no increase in the activity of the troublesome $^{222}$Rn background after calibration. Alpha emitters are also distributed throughout the detector and facilitate calibration of its response to $^{222}$Rn. Using the delayed coincidence of $^{220}$Rn-$^{216}$Po, we map for the first time the convective motion of particles in the XENON100 detector. Additionally, we make a competitive measurement of the half-life of $^{212}$Po, $t_{1/2}=(293.9\pm (1.0)_\text{stat}\pm (0.6)_\text{sys})\1{ns}$.
\end{abstract}


\maketitle

\section{Introduction}

Significant experimental progress in particle physics continues to be made in searches for rare events such as neutrinoless double-beta decay~\cite{Ostrovskiy:2016uyx} or scattering of dark matter particles~\cite{Undagoitia:2015gya}. Experiments that use the liquid noble elements xenon or argon are at the forefront of these searches~\cite{Albert:2014awa,Aprile:2013doa,Akerib:2015rjg,Amaudruz:2014nsa,Agnes:2015ftt,Tan:2016zwf}. As these detectors are scaled up to improve their sensitivity, the self-shielding of external radioactive sources yields lower backgrounds and thus further improvement in detector sensitivity. However, this feature renders calibration with external sources impractical and necessitates the development of novel strategies. 

Several aspects of the detector sensitivity need to be calibrated. The position-dependent energy response to both signal and background events and a detector's ability to discriminate between electronic and nuclear recoil events are required. The former requires mono-energetic lines, whereas the latter requires a broad energy spectrum of events. Any additional information that can be extracted from calibration could prove useful to improve the understanding of an existing detector and to develop future experiments.

Radioactive calibration sources that can be directly mixed into the liquid target promise to provide an alternate method to external sources for calibration of current and future detectors. A dissolved $^{83m}$Kr source was proposed in~\cite{Manalaysay:20091003,Kastens:20100510,Hannen:20111019} and employed in dark matter detectors~\cite{Agnes:2014bvk,Akerib:2015rjg} for calibration of the electronic recoil energy scale at low energies. Tritiated methane was also employed in LUX~\cite{Akerib:2015wdi} in order to exploit the beta decays that fall within the low energy range of interest for dark matter investigations.

Here, we use the XENON100 dark matter experiment~\cite{Aprile:2011dd} to characterize a radioactive source of dissolved $^{220}$Rn~\cite{Lang:2016zde} for use in current and future low-background experiments. The source is well suited to calibrate the low-energy electronic recoil background, $(2-30)\1{keV}$. In addition, the isotopes in the $^{220}$Rn decay chain provide alpha and beta radiation that improve our understanding of intrinsic $^{222}$Rn~\cite{Albert:2015vma,Weber:phdthesis}, which is a dominant source of background in dark matter experiments~\cite{Aprile:2015uzo, Akerib:2015cja, Amaudruz:2012hr, Araujo:2011as}, such that we can tag the daughter $^{214}$Pb beta event. Finally, given the short decay time of the whole decay chain, all introduced activity vanishes within one week, independent of a detector's volume or purification speed. In this way, $^{220}$Rn starkly contrasts tritiated methane, which must be proactively extracted from a detector with a high-temperature zirconium gas purifier in a xenon purification loop~\cite{Akerib:2015rjg} and, thereby, necessitates greater efforts to fully circulate and purify a detector of greater volume.

\section{The XENON100 Detector}\label{sec:detector}

The XENON100 detector, described in detail in~\cite{Aprile:2011dd}, is a cylindrical liquid/gas time projection chamber (TPC) that is $30\1{cm}$ in height and diameter and uses $62\1{kg}$ of high-purity liquid xenon as a dark matter target and detection medium. An energy deposition in the TPC produces scintillation photons and ionization electrons. The photons provide the prompt scintillation signal ($S1$). For the measurements presented here, the cathode, which is biased at $-12$\,kV and positioned at the bottom of the TPC ($Z=-300\1{mm}$), is combined with a grounded gate and an anode, biased at 4.4 kV and placed near the liquid-gas interface ($Z=0\1{mm}$), to define an electric field of $400\1{V/cm}$ in the liquid volume. This field drifts the electrons from the interaction site to the liquid-gas interface, where an $8.8\1{kV/cm}$ field extracts the electrons into the gas phase. Then, a second signal ($S2$) is generated through proportional scintillation. Both $S1$ and $S2$, measured in photoelectrons (PE), are observed by two arrays of photo-multiplier tubes (PMTs), one in liquid xenon at the bottom of the TPC below the cathode and the other in the gaseous xenon above the anode. The TPC is surrounded by a veto region containing $99\1{kg}$ of liquid xenon. For the data reported here, this volume was not instrumented and thus only serves as a passive volume.

A diving bell is used to keep the liquid level constant between the gate and anode meshes. Gaseous xenon is continuously recirculated at $2.6$~s.l.m. through a purification loop. The returning gas pressurizes the diving bell to approximately $2.1$~atm, with most of the gas pressure being relieved in the veto through a pipe that is used to define the height of the liquid level. In a separate loop, gaseous xenon is liquefied and returned to the top of the detector. This open design results in a vertical temperature gradient of $0.8\1{K}$ over the height of the TPC.

\section{$^{220}$R\MakeLowercase{n} Decay Chain and Observed Time Evolution}\label{sec:timeevo}

We present the results from a calibration campaign using a $33.6\1{kBq}$ $^{220}$Rn source. The suitability of this source for its employment under low-background conditions was previously reported in~\cite{Lang:2016zde}. The source contains $^{228}$Th electrolytically deposited on a stainless steel disc $30\1{mm}$ in diameter and housed in a standard vacuum vessel that is connected to the xenon gas purification system using 1/4" VCR piping. The $^{220}$Rn atoms emanate from the source and are flushed into the TPC through the xenon gas stream. With approximately $20\1{m}$ of piping and a ($40\pm10)$\% source emanation efficiency, we estimate that we acquire $(1.8\pm0.5)\times 10^{9}$ $^{220}$Rn atoms in the entire detector while the source is open for $1.7\1{days}$. A total of $1.7\times 10^{7}$ decays are observed in the active region during the full calibration run, while the remainder is in the veto region.

\begin{figure}[!t] 
\centering
\includegraphics[width=1\columnwidth,trim={8mm 0 8mm 25mm},clip=true]{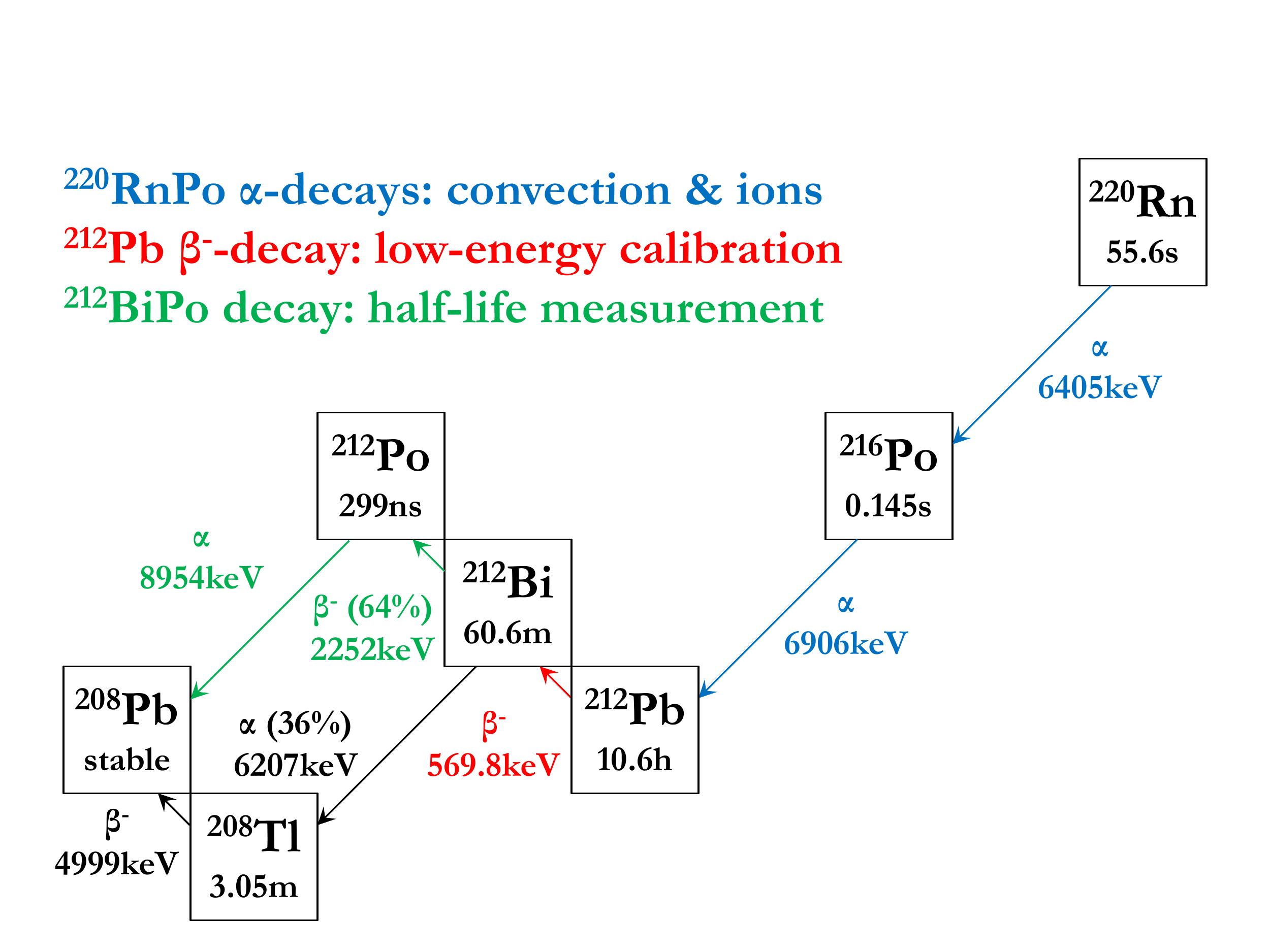}
\caption{The decay chain of $^{220}$Rn, following its emanation from a $^{228}$Th source. The three main analyses presented in this paper are labeled in color. $Q$-values are taken from~\cite{Qvalues}, and half-lives and branching ratios from~\cite{NuclearData}.}
\label{fig:DecayChain}
\end{figure}

\begin{figure}[!htb] 
\centering
\includegraphics[width=0.97\columnwidth]{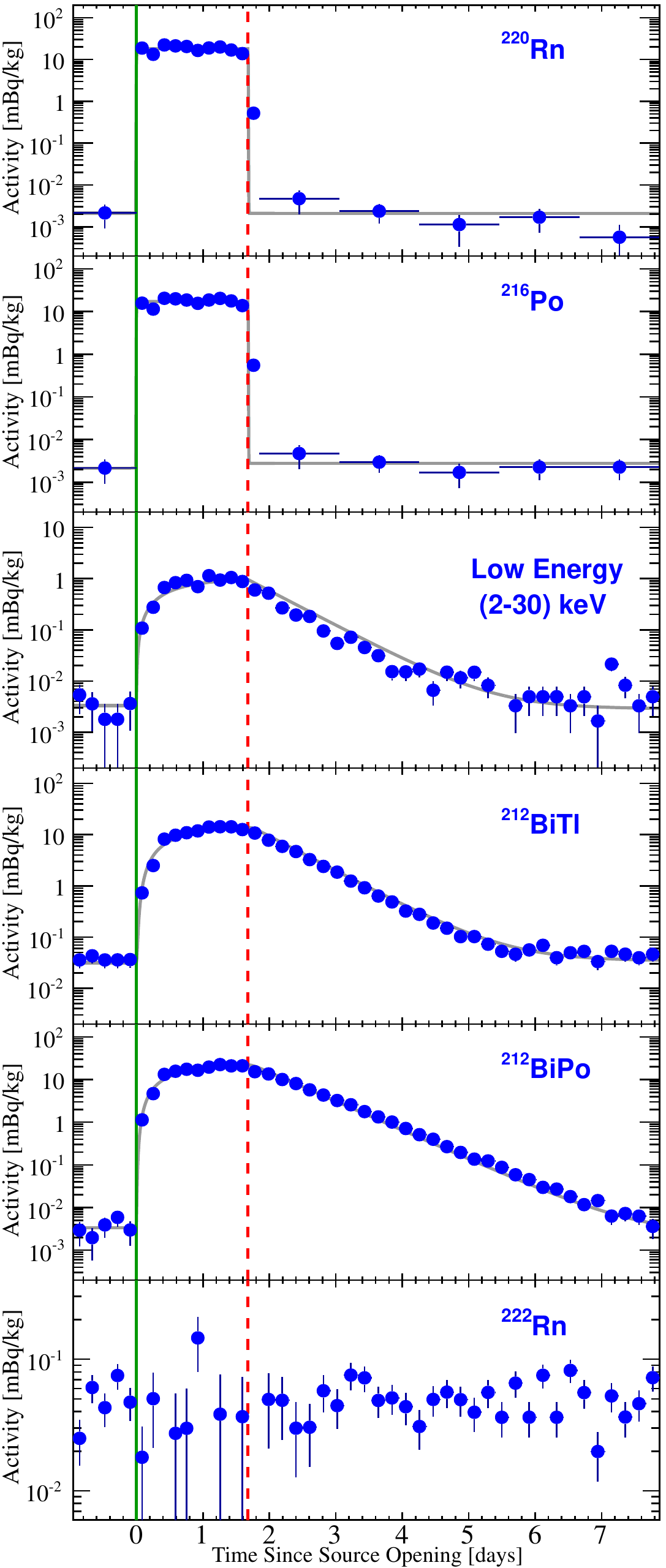}
\caption{Time evolution of isotopes in the $^{220}$Rn decay chain and the $^{222}$Rn background. Times at which the source was opened and closed are indicated by the solid green and the dashed red lines. The opening of the source is defined as time $t=0$. Gray curves show the numerical model of the growth and decay of each isotope, scaled to the observed rate. Binning for $^{220}$Rn and $^{216}$Po is adjusted for times of low activity.}
\label{fig:TimeEvolution}
\end{figure}

The relevant portion of the $^{220}$Rn decay chain is shown in Figure~\ref{fig:DecayChain}. We first address the overall viability of this calibration source and defer the description of the respective event selection criteria to the following sections, which also detail the physics that can be extracted from each of the steps in the decay chain. Figure~\ref{fig:TimeEvolution} shows the isotopic temporal evolution as observed in XENON100. These rates are corrected for deadtime effects, which arise from significant DAQ saturation due to a trigger rate of $\mathcal{O}(100)\1{Hz}$ while the source is open. Deadtime information is recorded once per minute.

The short-lived isotopes, $^{220}$Rn ($t_{1/2}=55.6\1{s}$) and $^{216}$Po ($t_{1/2}=0.145\1{s}$), grow into the active region of the detector within minutes after opening the source, and they quickly decay once the source is closed. Delayed coincidence of $^{220}$Rn and $^{216}$Po provides the means to detail fluid dynamics within the detector volume. The flow pattern of particles is particularly interesting because it has the potential to improve the efficiency of purification systems through the identification of dead regions. Furthermore, the pattern may inspire new methods to identify and reduce the $^{222}$Rn background by capitalizing on the sequence of the $^{220}$Rn decay chain. Due to the one-minute intervals in which deadtime information is recorded, no features are visible in the rise of $^{220}$Rn or $^{216}$Po activity.

The primary utility of a $^{220}$Rn calibration source comes from the ground-state beta decay of $^{212}$Pb with a $Q$-value of $569.8\1{keV}$ and a branching ratio of 11.9\%. This decay results in low-energy electronic recoils $(2-30)\1{keV}$ that can be used for background calibration in dark matter searches. As we show here, the isotope's long half-life, $10.6\1{hours}$, gives it ample time to spread throughout the entire detector volume while being sufficiently short to allow the activity to decay within a week. We find that 3 in every $10^4$ decays are useful for low-energy electronic recoil calibration. For comparison, this ratio is two orders of magnitude higher than that of typical external gamma sources in XENON100, and it is expected to remain constant as detectors become larger.

The beta decays of $^{212}$Bi may be selected with high purity due to their delayed coincidence with the alpha decays of $^{212}$Po, occurring shortly afterward ($t_{1/2}=299\1{ns}$). These BiPo events provide the means to confirm that the introduced activity indeed spreads throughout the entire TPC (see Figure~\ref{fig:BiPoSpatialDist}). Additionally, they yield the most accurate measurement of the introduced activity's dissipation time of $\sim7\1{days}$, as seen in Figure~\ref{fig:TimeEvolution}.

Further utility of this source comes from the 2.6-MeV gamma decay of $^{208}$Tl, which is close to the 2.5-MeV double-beta decay of the $^{136}$Xe. Due to other low-energy gammas that accompany it, multiple steps are created in the energy spectrum and can be exploited in calibration. The alpha decays of $^{220}$Rn, $^{216}$Po, and $^{212}$Bi can be used to calibrate position-dependent light and charge collection efficiencies at high energies (see Figure~\ref{fig:AlphaMap}).

Figure~\ref{fig:TimeEvolution} also shows the expected behavior of the various isotopes based on a simple calculation of the exponential decay chain. This treatment effectively assumes instantaneous and complete mixing of all isotopes. As can be seen, this model provides an excellent description of the observed time evolution. A comparison of the short- and long-lived portions of the decay chain suggests that there are more low-energy events than expected from long-lived isotopes.

Gas routing in XENON100 causes most of the activity to be pumped and retained in the veto. A GEANT4 Monte Carlo simulation of the XENON100 detector geometry demonstrates that the probability a gamma decay of $^{212}$Pb ($^{212}$Bi; $^{208}$Tl) in the veto region induces a low-energy single scatter event in the $34\1{kg}$ fiducial volume used in~\cite{Aprile:2012nq} is $6\times10^{-6}$ $ (3\times10^{-5}; 2\times10^{-4})$. We estimate that there are 1000 of these decays in the veto for every true $^{212}$Pb decay in the $34\1{kg}$ fiducial volume. It then follows that for every 0.012 true low-energy $^{212}$Pb decays there are 0.236 events that result from the gamma decays of long-lived isotopes in the veto. Therefore, we conclude that $5\%$ of the low-energy events that fall within the fiducial volume are truly caused by the low-energy beta decays of $^{212}$Pb. By comparison, the number of $^{212}$BiPo events $(1.6\times 10^{5})$, which is $\sim2/13$ of the total activity of the decay chain, shows that only 6\% $(1.0\times 10^{6})$ of the total number of observed events actually originate in the TPC.

\section{Alpha Spectroscopy}\label{sec:spectroscopy}

\begin{figure}[!tb] 
\centering
\includegraphics[width=1.00\columnwidth]{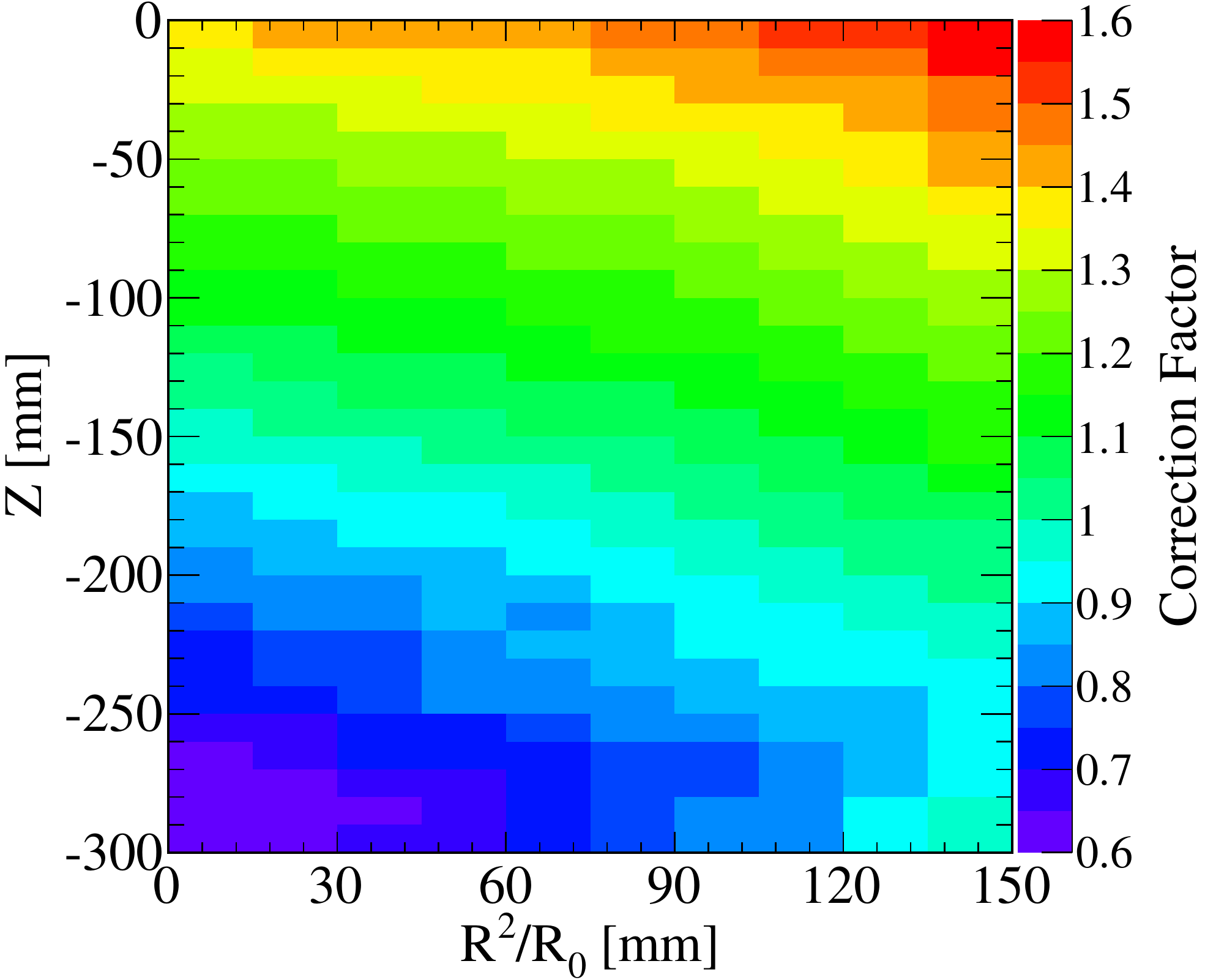}
\caption{Light correction map of XENON100 for high-energy alpha events, generated from $^{220}$Rn and $^{216}$Po decays. The radial parameter is defined according to the detector's radius, $R_0=153\1{mm}$. Events at low (high) $Z$ and low (high) radius~$R$ have the highest (lowest) light collection and thus a correction factor less (greater) than unity.}
\label{fig:AlphaMap}
\end{figure}

The interactions of alpha particles in liquid xenon are easily identifiable because they produce tracks with a large ionization density, which results in small $S2$'s and large $S1$'s compared to electronic and nuclear recoils. This difference is due to a higher probability of recombination and allows the alphas to be selected based on their $S1$ yield, thereby rejecting backgrounds from beta or gamma sources. 

We derive a correction factor to account for variations in light collection efficiency across the TPC~\cite{Aprile:2011dd}. As the PMT bases in XENON100 have been optimized for low-energy events in the search for dark matter, a dedicated correction map is required to account for nonlinearity in the response of PMTs to high-energy events~\cite{Aprile:2015uzo}. For each spatial bin shown in Figure~\ref{fig:AlphaMap}, the arithmetic average of the two observed mean scintillation values of the 6.4- and 6.9-MeV alpha decays of $^{220}$Rn and $^{216}$Po is calculated. Then, these values are scaled by the volume-averaged value to obtain a relative correction factor, shown in Figure~\ref{fig:AlphaMap}. The radial parameter $R^2/R_0$ is defined according to the detector's radius, $R_0=153\1{mm}$. Events at low (high) $Z$ and low (high) $R$ have the highest (lowest) light collection and thus a correction factor less (greater) than unity.

\begin{figure}[!tb] 
\centering
\includegraphics[width=1.00\columnwidth]{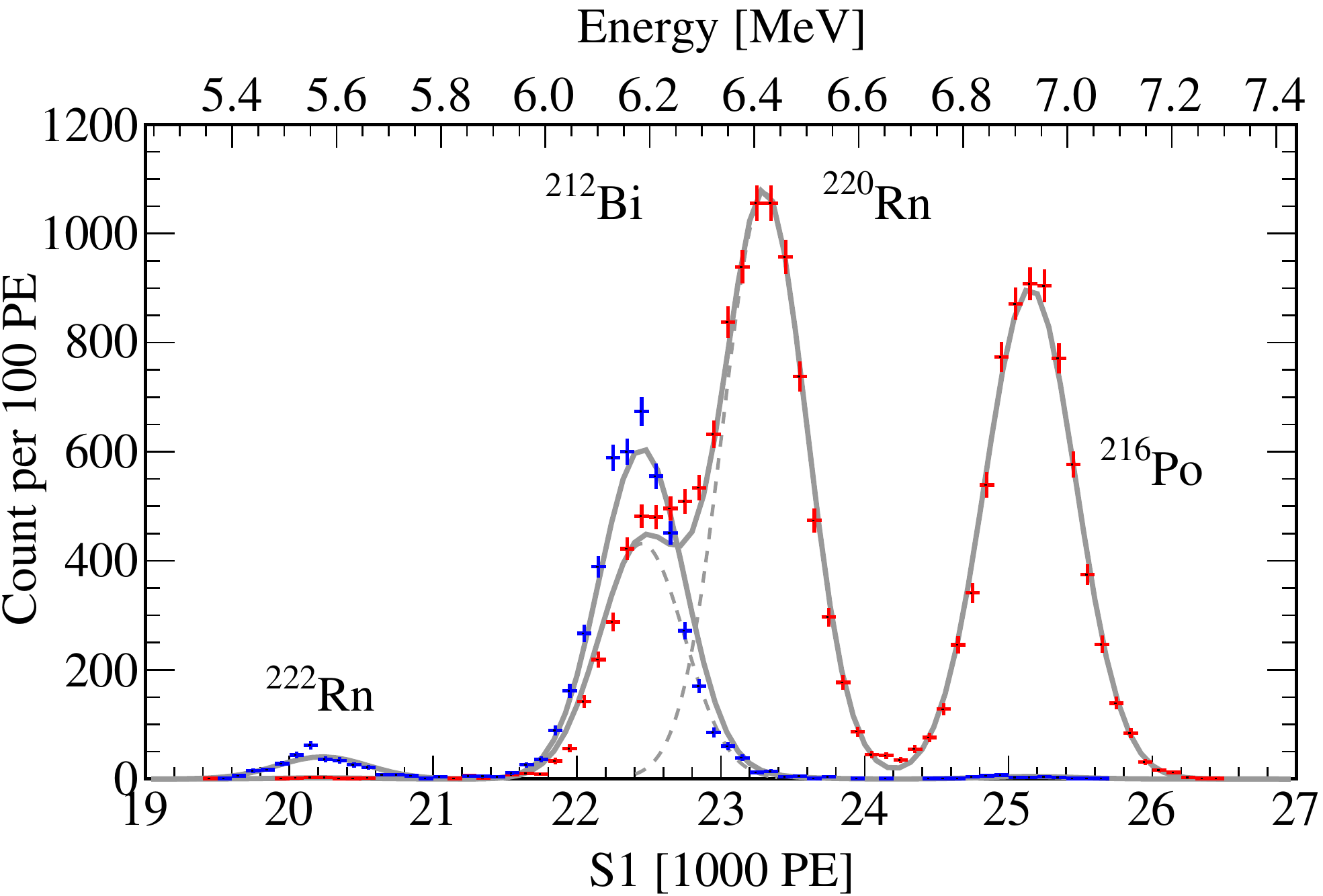}
\caption{The alpha spectrum of the $^{220}$Rn decay chain is shown integrated over times during which the source is open ($40.3\1{hours}$, red) and after ($148.7\1{hours}$, blue) the source is closed. The constant background of $^{222}$Rn (Figure~\ref{fig:TimeEvolution}) is visible only over the longer time period.}
\label{fig:AlphaSpectroscopy}
\end{figure}

The energy spectrum of the alpha decays of $^{222}$Rn, $^{212}$Bi, $^{220}$Rn, and $^{216}$Po is shown in Figure~\ref{fig:AlphaSpectroscopy}, after applying the alpha light correction map. The population of alphas is split into periods during which the source is open (red) and after the source is closed (blue). These events have been selected in the fiducial volume defined by $R\leq 100\1{mm}$ and $-200\1{mm}\leq Z\leq-5\1{mm}$ in order to avoid the degradation in energy resolution near the cathode and thereby optimize the identification of different isotopes. The energy, mean scintillation signal, and light yield (LY) are listed for each isotope in Table~\ref{table:Alphas}. The light yield is constant in this energy range to within 0.3\%. Our ability to identify and characterize the various alpha particles is the foundation of the multiple modes of delayed coincidence that are discussed in subsequent sections.

\begin{table}[!htb]
\centering
\begin{tabular}{c c l c c}
\hline\hline
Isotope & \: $Q$ [MeV] \: & \: \;\;S1 [PE] \: & \: $\sigma$ [PE] \: & \: LY [PE/keV] \\\hline
$^{222}$Rn & 5.590 & \;20231 $\pm$ 17 & 315 & 3.62 \\
$^{212}$Bi & 6.207 & \;22451 $\pm$ 3  & 296 & 3.62 \\
$^{220}$Rn & 6.405 & \;23306 $\pm$ 3  & 291 & 3.64 \\
$^{216}$Po & 6.906 & \;25152 $\pm$ 4  & 316 & 3.64 \\
\hline\hline
\end{tabular}
\caption{$Q$-values, scintillation values (means and widths at $400\1{V/cm}$), and calculated light yields of each alpha decay in Figure~\ref{fig:AlphaSpectroscopy}.}
\label{table:Alphas}
\end{table}

\section{$^{220}$R\MakeLowercase{n}-$^{216}$P\MakeLowercase{o} Coincidence and Convection}
\label{sec:RnPo}

The combination of spatial and temporal information permits us to match $^{216}$Po with its parent $^{220}$Rn. As a result, we measure the position resolution at high energies, map the fluid dynamics, and calculate a lower limit on the drift speed of $^{216}$Po ions in the XENON100 TPC.

We select $^{220}$Rn$^{216}$Po (RnPo) pairs within $3\sigma$ of the respective scintillation peaks (see Figure~\ref{fig:AlphaSpectroscopy}) with the requirement that a candidate $^{216}$Po decay occur within $1\1{s}$ and $8\1{mm}$ of a candidate $^{220}$Rn parent. The time condition selects 99\% of all pairs, and the spatial condition prevents the formation of pairs in which polonium is too distant to be causally related to radon. Under these conditions, less than 0.3\% of the $^{220}$Rn candidates have more than one possible $^{216}$Po partner, and consequently they are removed from the analysis. A total of 45441 RnPo pairs is found in multiple calibrations between June and November 2015.

Each RnPo pair provides differential position values in the vertical ($\Delta Z$) and the horizontal ($\sqrt{(\Delta X)^2+(\Delta Y)^2}$) directions. The resultant distributions yield upper limits on the position resolutions at high energies: $\sigma_Z=0.2\1{mm}$ and $\sigma_{XY}=0.7\1{mm}$. These resolution limits are better than the resolutions reported at low energies in~\cite{Aprile:2011dd} ($\sigma_Z=0.3\1{mm}$, $\sigma_{XY}=3\1{mm}$) due to the significantly larger signals.

Moreover, we study fluid dynamics of the liquid xenon using RnPo pairs. To fully appreciate the features of bulk atomic motion in the XENON100 TPC, we scan the full range of azimuthal and vertical angles, ultimately selecting a rotated view of the cylindrical TPC from the side at $\phi=-45^\circ,\theta=90^\circ$, with appropriate coordinate transformations. In a projection of all events on a (Cartesian) cross sectional (i.e. the $YZ$-plane), the density of events appears distorted. We therefore introduce the parameter $\tilde{Y}$, derived in the Appendix, which preserves uniformity in number density in a projection of the cylindrical TPC onto a plane containing its central axis:
\begin{equation}\label{eq:Ytilde}
\tilde{Y} = R_0\bigg[\frac{2}{\pi}\bigg(u-\frac{1}{2}\sin(2u)\bigg)-1\bigg]\ ,
\end{equation}
where
\begin{equation}
\label{eq:theta}
u = \arccos\bigg(-\frac{Y'}{R_0}\bigg)\ ,
\end{equation}
$Y' = -X\sin(\phi)+Y\cos(\phi)$ is the relevant rotated coordinate, and $R_0=153\1{mm}$ is the radius of the TPC.

\begin{figure*}\centering
        {\includegraphics[width=0.49\textwidth]{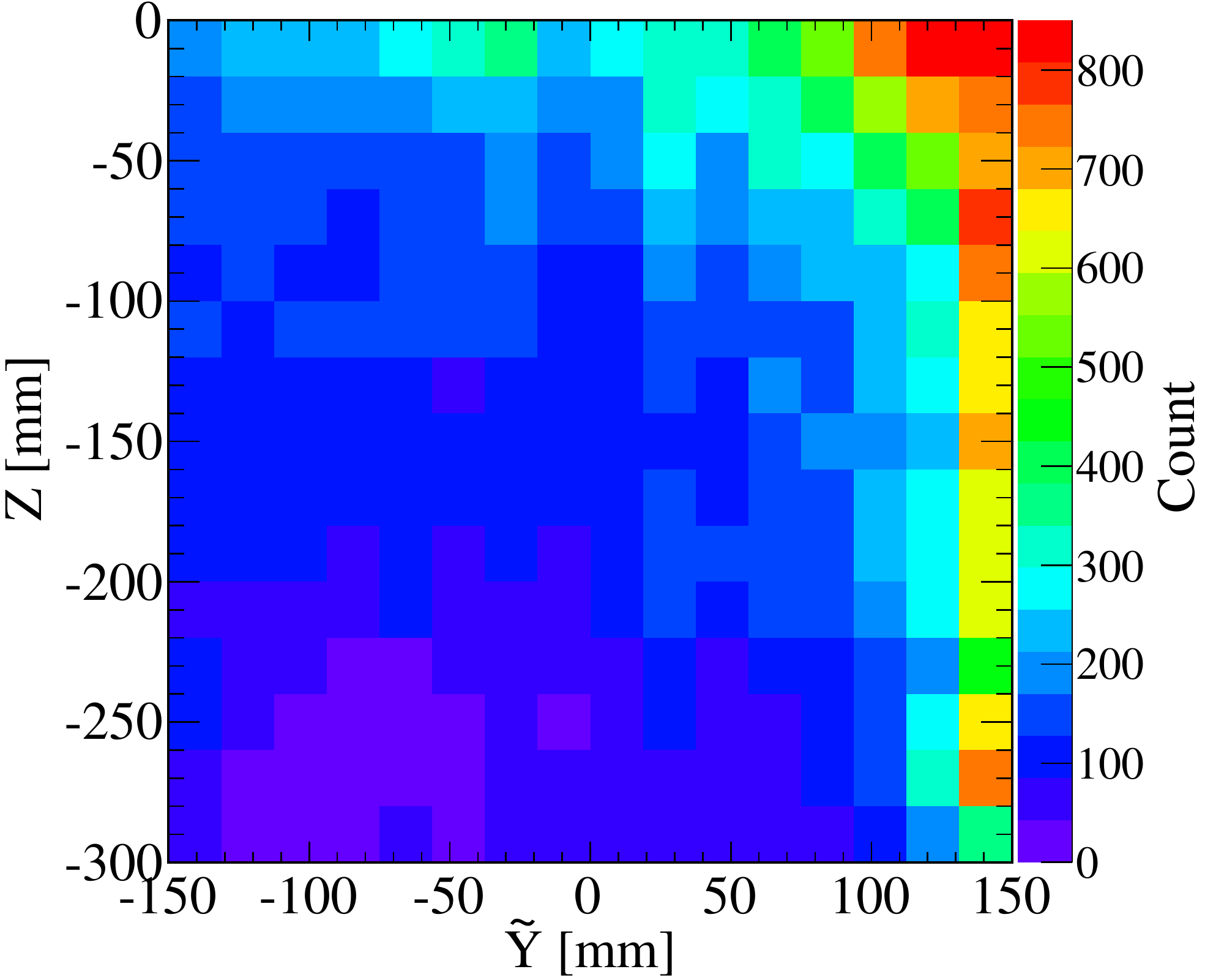}}
    \hfill
         {\includegraphics[width=0.49\textwidth]{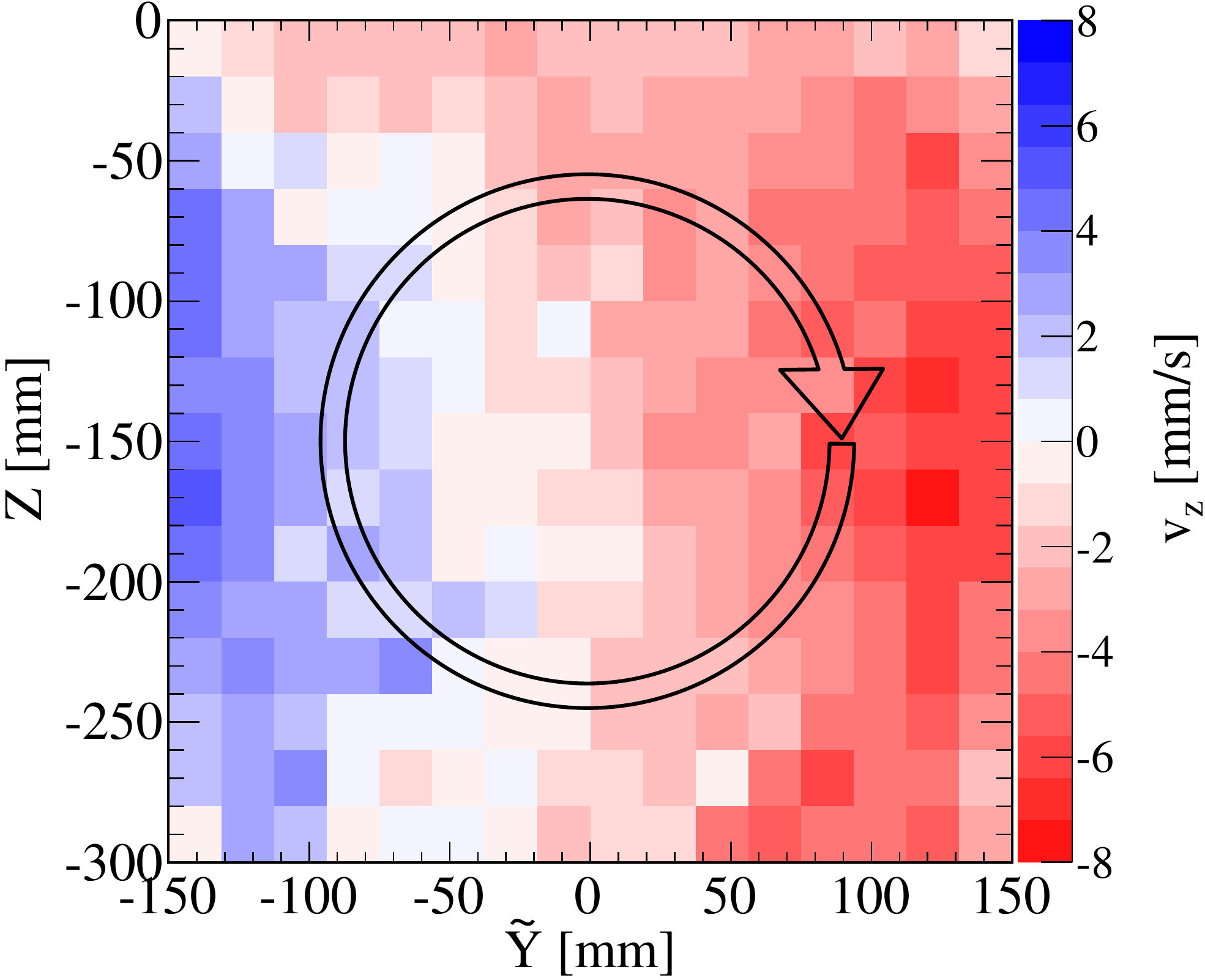}}

         {\includegraphics[width=0.49\textwidth]{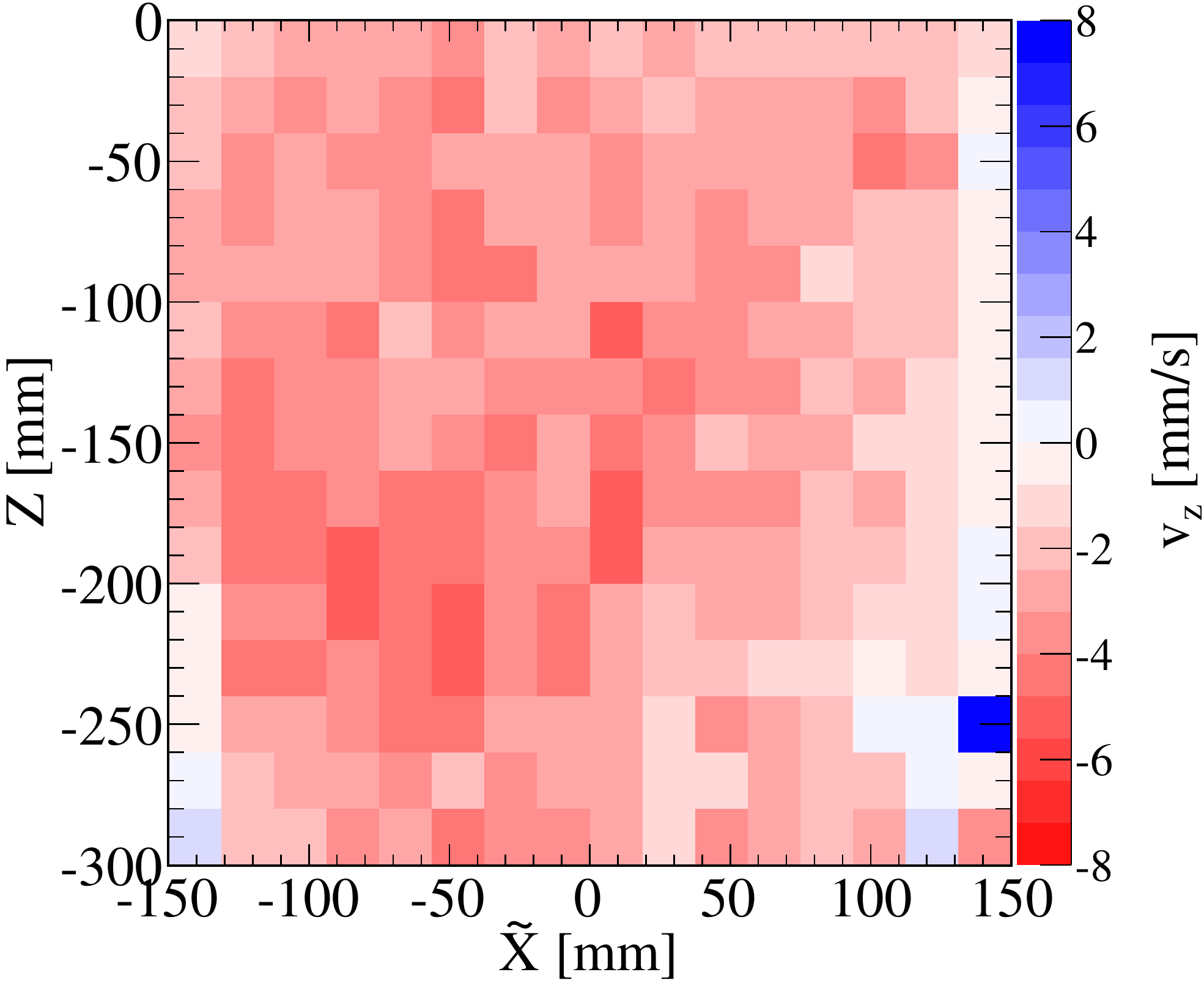}}
    \hfill
        {\includegraphics[width=0.49\textwidth]{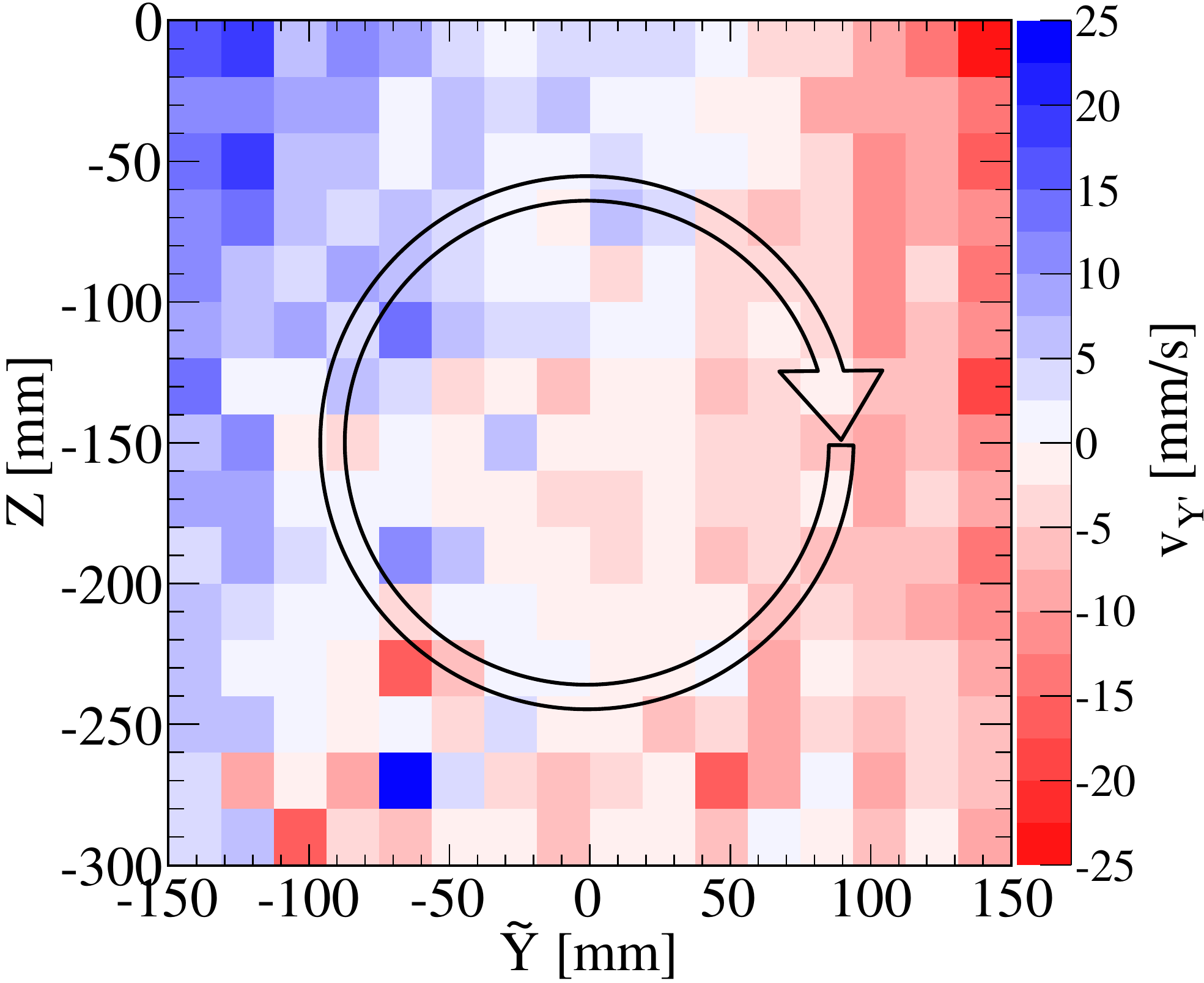}}
\caption[]
        {(Top left) Number density of $^{220}$Rn$^{216}$Po pairs across the TPC. Due to the short half-lives at the beginning of the decay chain, these events are still concentrated near the top of the TPC around $\tilde{Y}=153\1{mm}$. (Top right) Delayed coincidence of these pairs is used to map the average $z$-component of atomic velocity as a function of position $\tilde{Y}$. As indicated by the arrow, atomic motion in XENON100 primarily results in a single convection cell, viewed here along the direction of its angular momentum vector. (Bottom left) The average $z$-component of atomic velocity is also shown as a function of position $\tilde{X}$. In this view, one predominantly sees downward motion, as expected from the substantially larger number of atoms that move downward rather than upward. (Bottom right) Similarly, we find evidence of convective motion when we determine the average $Y'$-component of atomic velocity as a function of position $\tilde{Y}$. However, we note that the observed range of horizontal speeds is inflated compared to the range of vertical speeds due to the superior resolution of the vertical position.}
    \label{fig:Convection}
    \end{figure*}

In Figure~\ref{fig:Convection}~(top left), the number density of RnPo pairs is shown as previously selected. The pairs tend to concentrate along the outer surface near $\tilde{Y}=153\1{mm}$. A large gradient in the number density exists because the half-lives of both $^{220}$Rn and $^{216}$Po are much shorter than the time it takes an atom to fully traverse the TPC.

In Figure~\ref{fig:Convection}~(top right), we show the average $z$-velocities of the $^{216}$Po daughter, $v_Z=\Delta Z/\Delta t$, as a function of position $\tilde{Y}$, where $\Delta t$ is the time the $^{216}$Po atom takes to decay. For $\tilde{Y}>0$ ($\tilde{Y}<0$), particles move downward (upward) at speeds up to $7.2\1{mm/s}$ ($4.8\1{mm/s}$). Ergo, the TPC of XENON100 is a single convection cell whose net angular momentum lies along $\phi\approx135^\circ$. This pattern is observed to be unchanged between two separate calibration campaigns taken five months apart.

In Figure~\ref{fig:Convection}~(bottom left), we determine the average $z$-velocities of $^{216}$Po in terms of the partner coordinate $\tilde{X}$, which is calculated following Eqs.~\ref{eq:Ytilde} and~\ref{eq:theta} with $X' = X\cos(\phi)+Y\sin(\phi)$. As one would expect in this view, due to the larger number of atoms that move downward rather than upward, one predominantly sees downward motion throughout the TPC.

Moreover, in Figure~\ref{fig:Convection}~(bottom right), we find additional evidence of convection showing the average $Y'$-component of atomic velocity, $v_{Y'}=\Delta Y'/\Delta t$, as a function of position $\tilde{Y}$. We note, however, that the observed range of horizontal speeds is inflated compared to the range of vertical speeds due to the superior resolution of the vertical position.

Such convective motion, which is most likely driven by the xenon recirculation flow, holds significant implications for the deployment of calibration sources and for the development of techniques for background mitigation in future experiments. The $^{214}$Pb daughter is the main contributor to the low-energy electronic recoil background that arises from the $^{222}$Rn decay chain. A known convection pattern can be used to track this $^{214}$Pb ($t_{1/2}=27.1\1{min}$) from the site of its parent $^{218}$Po's alpha decay ($t_{1/2}=3.1\1{min}$) or to track its daughter $^{214}$Bi ($t_{1/2}=19.9\1{min}$). Consequently, a low-energy $^{214}$Pb decay could be tagged, effectively reducing the electronic recoil background.

Finally, due to the difference between the minimum and maximum velocities in Figure~\ref{fig:Convection}, we conclude that there is a subdominant contribution to the total particle motion that results from the electric field ($400\1{V/cm}$) applied to $^{216}$Po ions. As stated previously, Figure~\ref{fig:Convection} shows a concentration of RnPo pairs in the region of downward velocities because the atoms decay on time scales shorter than the time required to traverse the TPC. Simply averaging the pairs' position-dependent velocities over the full TPC volume would inevitably introduce a downward bias in any measurement of ion drift. We thus avoid this subtle bias by taking two distinct sets, one of upward and one of downward atoms, in $|\tilde{Y}|>100\1{mm}$ to calculate the variance-weighted mean velocities, $\bar{v}_{\mathrm{up}}$ and $\bar{v}_{\mathrm{down}}$, on either side of the TPC for each of ten equal $Z$ bins in the range $[-200,-100]\1{mm}$. Then, the velocity offset within each $Z$ bin is found by averaging the two components: $v_\text{offset} = (\bar{v}_{\mathrm{up}}+\bar{v}_{\mathrm{down}})/2$. Consequently, the mean offset over all ten bins is found to be $\bar{v}_\text{offset}=(0.9\pm0.3)\1{mm/s}$ toward the cathode. This offset constitutes a lower limit on the ion drift speed of $^{216}$Po, corresponding to a completely (positively) ionized population. Our limit is consistent with the distribution of ion drift speeds presented by EXO-200 for $^{218}$Po at $380\1{V/cm}$~\cite{Albert:2015vma}.

Delayed coincidence of $^{212}$Bi and $^{208}$Tl (BiTl) is also attempted following a methodology similar to that of RnPo. With atomic speeds up to $\sim7\1{mm/s}$ in XENON100, the $^{208}$Tl atom can travel up to $\sim140\1{cm}$ during its $3\1{minute}$ half-life. This distance is much larger than the $30\1{cm}$ dimension of XENON100, making it impossible to accurately match BiTl pairs. However, BiTl coincidence may be a useful component of this calibration source for meter-scale detectors such as XENON1T or single-phase detectors such as EXO-200.

\section{The half-life of $^{212}$P\MakeLowercase{o}}
\label{sec:bipo}

\begin{figure}[!tb] 
\centering
\includegraphics[width=1.00\columnwidth]{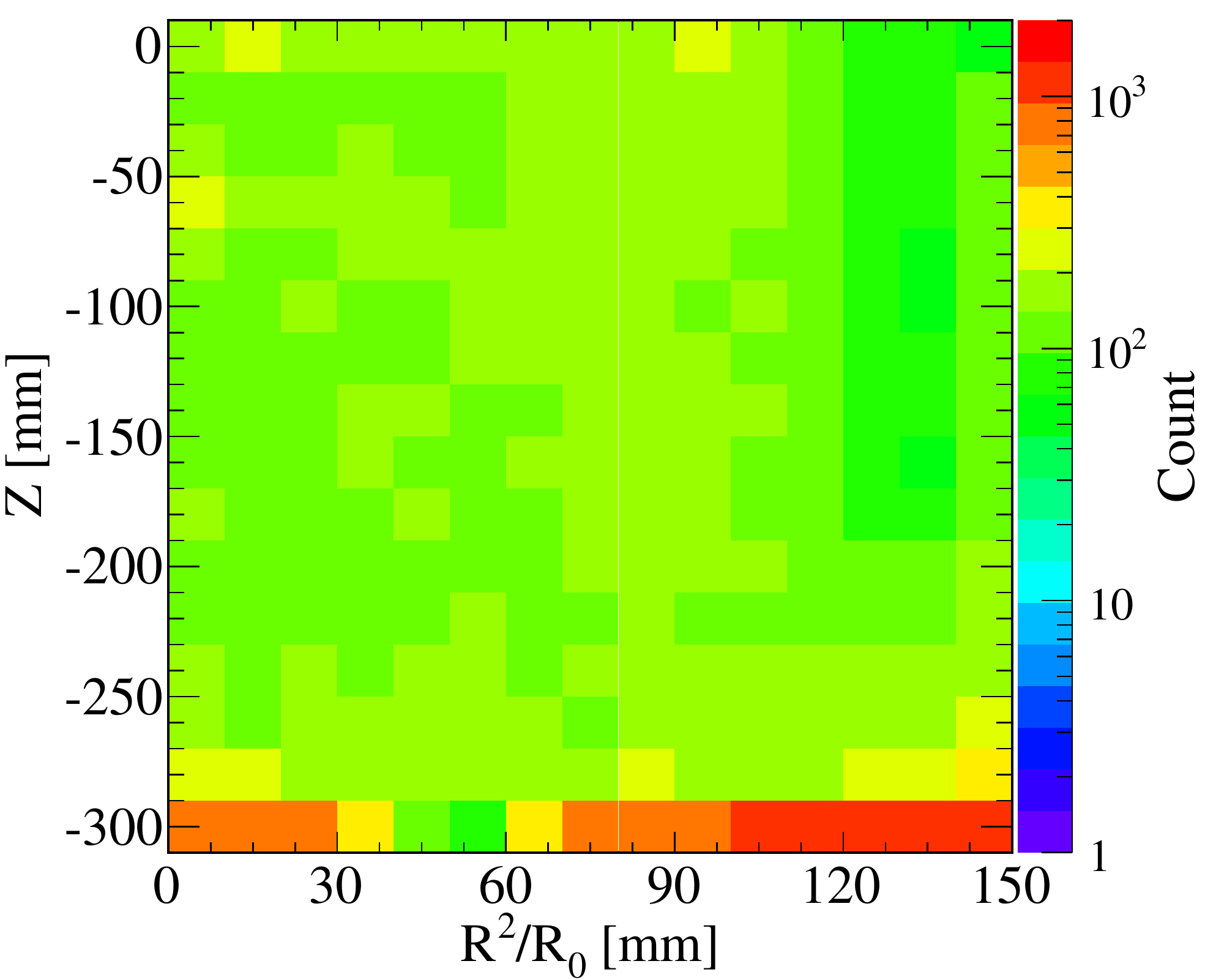}
\caption{The spatial distribution of low-energy $^{212}$BiPo decays. The events permeate the entire active region of the TPC.}
\label{fig:BiPoSpatialDist}
\end{figure}

The beta decay of $^{212}$Bi and the alpha decay of $^{212}$Po are easy to identify because they occur in quick succession within a single acquisition time window. The selection of these BiPo events is made within the range $20\1{PE}<S1_{\beta}<7000\1{PE}$, $10000\1{PE}<S1_{\alpha}<55000\1{PE}$, and with the requirement that the $S1_{\beta}$ appear in the waveform before the $S1_{\alpha}$. The S2 signals from the two decays overlap in time and thus are not used in the analysis. Figure~\ref{fig:BiPoSpatialDist} shows that the BiPo events are distributed throughout the active region, with significant clustering at the cathode.

\begin{figure}[!htb] 
\centering
\includegraphics[width=1.00\columnwidth,trim={0 0 0 0},clip=true]{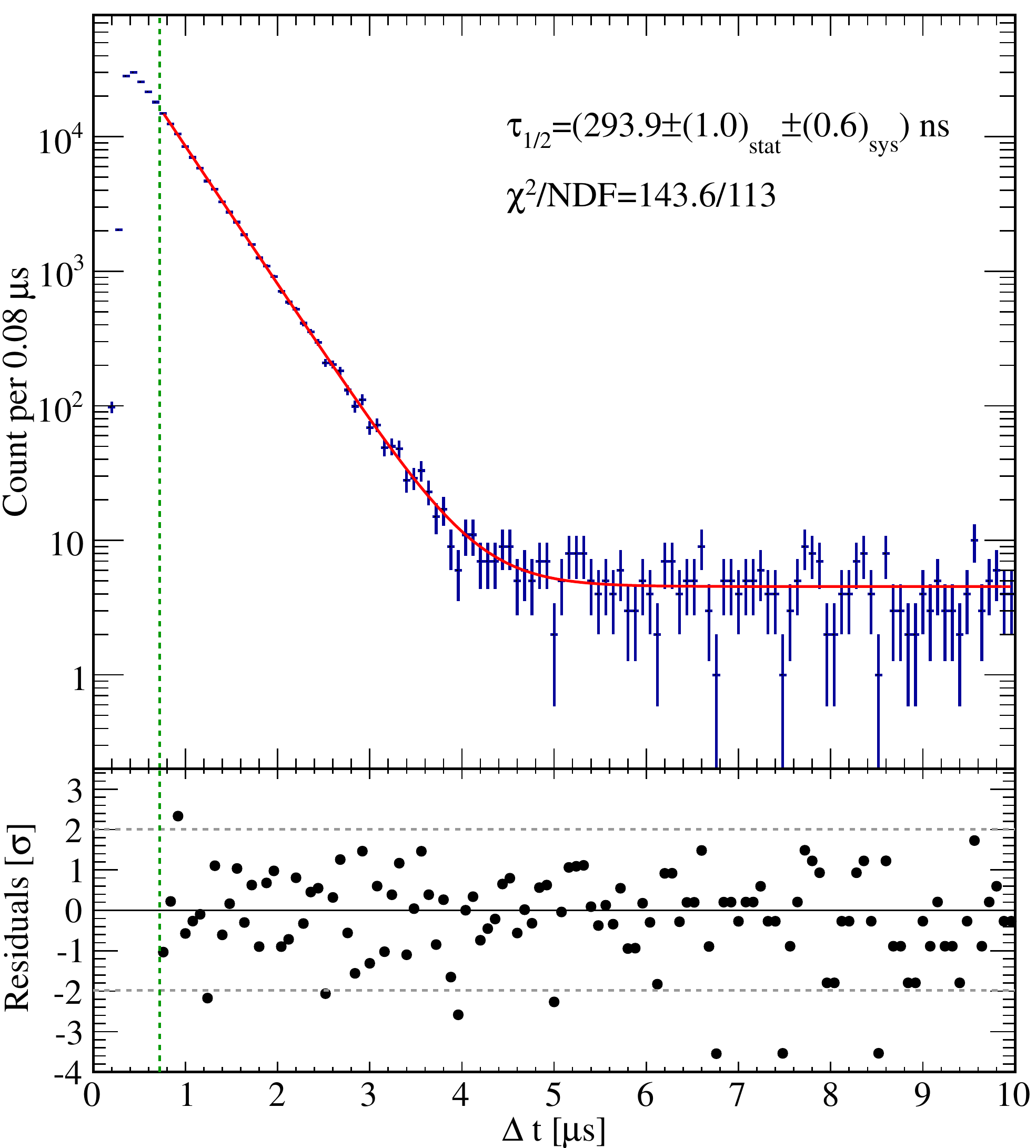}
\caption{(Top) Time difference of BiPo coincidence events together with a fit of the half-life of $^{212}$Po. The lower bound of the fit is set at $720\1{ns}$ to optimize the combined statistical and systematic uncertainty. (Bottom) The residuals of the fit in the top panel.}
\label{fig:HalfLife}
\end{figure}

To ensure that the S1 of the alpha decay is identified by the peak finder, it must come before the S2 of the beta decay. Hence, the drift time of the ionization electrons from the beta decay of $^{212}$Bi must exceed the decay time of the $^{212}$Po alpha decay. We consequently exclude events within $5\1{mm}$ of the liquid surface. Figure~\ref{fig:HalfLife} shows the resulting distribution of time differences of the selected alpha and beta decays, calculated from the differences of their respective S1 peaks. We infer from this distribution that this BiPo sample is $>99.75$\% pure.

The distribution of $^{212}$Po decay times is fitted with an exponential model of the event rate, $N(\Delta t)=N_0e^{-\Delta t/\tau}+B$, where $N_0$ is the event rate at small time differences, $\tau$ is the lifetime, and $B$ is the background rate. Fitting over the range $[0.72,10]\1{\mu s}$, we find a half-life $t_{1/2}=\tau\text{ln}(2)=(293.9\pm (1.0)_\text{stat})\1{ns}$. The residuals of this fit are shown in the bottom panel of Figure~\ref{fig:HalfLife}.

\begin{figure}[!htb] 
\centering
\includegraphics[width=1.00\columnwidth,trim={0 0 0 0},clip=true]{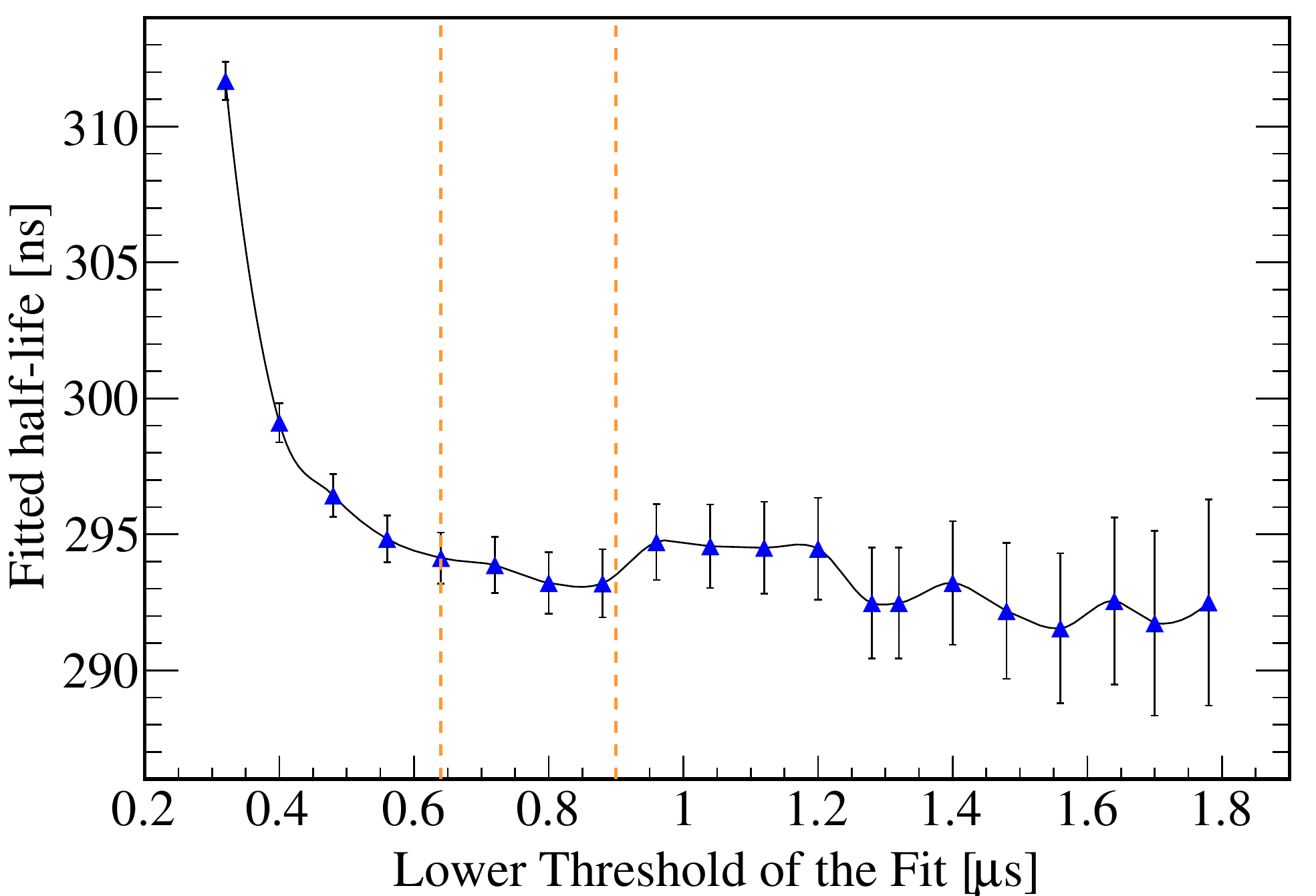}
\caption{The fitted half-life of $^{212}$Po as a function of the minimum time difference that is considered. The lower bound of the half-life measurement is set at $0.72\1{\mu s}$. Based on the range $[0.64,0.90]\1{\mu s}$, we estimate a 0.5-ns contribution to the total systematic uncertainty that results from the lower threshold of the time difference.}
\label{fig:LowerBound}
\end{figure}

Various systematic effects are considered for the uncertainty of this half-life measurement. The minimum time difference for the fit, $0.72\1{\mu s}$, is chosen to minimize the combined statistical and systematic uncertainty. Figure~\ref{fig:LowerBound} shows the fitted half-life as a function of this minimum time difference. The RMS value of the four points in the range $[0.64,0.90]\1{\mu s}$, $0.5\1{ns}$, is taken as the systematic uncertainty resulting from the fit range. The clock of the digitizer contributes less than $0.3\1{ns}$ to the systematic uncertainty. We repeat the measurement while varying the lower energy thresholds of our selection of $S1_{\alpha}$ and $S1_{\beta}$ in their respective ranges $[6000,12000]\1{PE}$ and $[20,500]\1{PE}$. From these variations, we estimate a $0.1\1{ns}$ contribution to the uncertainty. Furthermore, we find that the result is independent of the temporal resolution and the choice of binning. Jitter from finite ADC sampling, $S1$ scintillation properties, and other effects that modify the time stamp of individual $S1$ peaks average out. Ultimately, we measure the half-life of $^{212}$Po to be $t_{1/2}=(293.9\pm (1.0)_\text{stat}\pm (0.6)_\text{sys})$~ns. This measurement agrees with a recent half-life measurement from Borexino, $t_{1/2}=(294.7\pm(0.6)_\text{stat}\pm(0.8)_\text{sys})\1{ns}$ \cite{Bellini:2012qg}, but is slightly smaller than the most recent measurement from DAMA/LIBRA, $t_{1/2}=(298.8\pm(0.8)_\text{stat}\pm(1.4)_\text{sys})\1{ns}$~\cite{Belli2014}.

\section{Low-Energy Calibration}
\label{sec:lowenergy}

The decay of $^{212}$Pb emits beta particles that are effective for the calibration of liquid noble gas detectors in the search for dark matter. The direct decay of $^{212}$Pb to the ground state of $^{212}$Bi, occurring with a branching ratio of 11.9\%, is the relevant decay for low-energy electronic recoil calibration. Specifically, 10\% of these direct decays fall within the low-energy range of interest, $(2-30)\1{keV}$. The source is viable in this capacity as long as the activity of $^{212}$Pb spreads throughout the entire active region of the TPC.

In order to identify low-energy decays with high efficiency, three types of selection cuts are applied following previous XENON100 dark matter analyses of electronic recoils~\cite{Aprile:2014eoa, Aprile:2015ade, Aprile:2015ibr}. The first type includes cuts that remove events with excessive levels of electronic noise, ensuring that we choose actual particle interactions. The second type checks for consistency between the drift time and the width of the S2. Finally, we select the relevant low-energy beta decays by requiring a single scatter interaction with a scintillation signal in the range $3\1{PE}<S1<60\1{PE}$, which corresponds to the energy range $(2-30)\1{keV}$~\cite{Aprile:2014eoa, Aprile:2015ade}. However, as discussed in Section~\ref{sec:timeevo}, 95\% of these events are induced by gammas that travel into the TPC from decays in the veto region. 

\begin{figure}[!tb] 
\centering
\includegraphics[width=1.00\columnwidth]{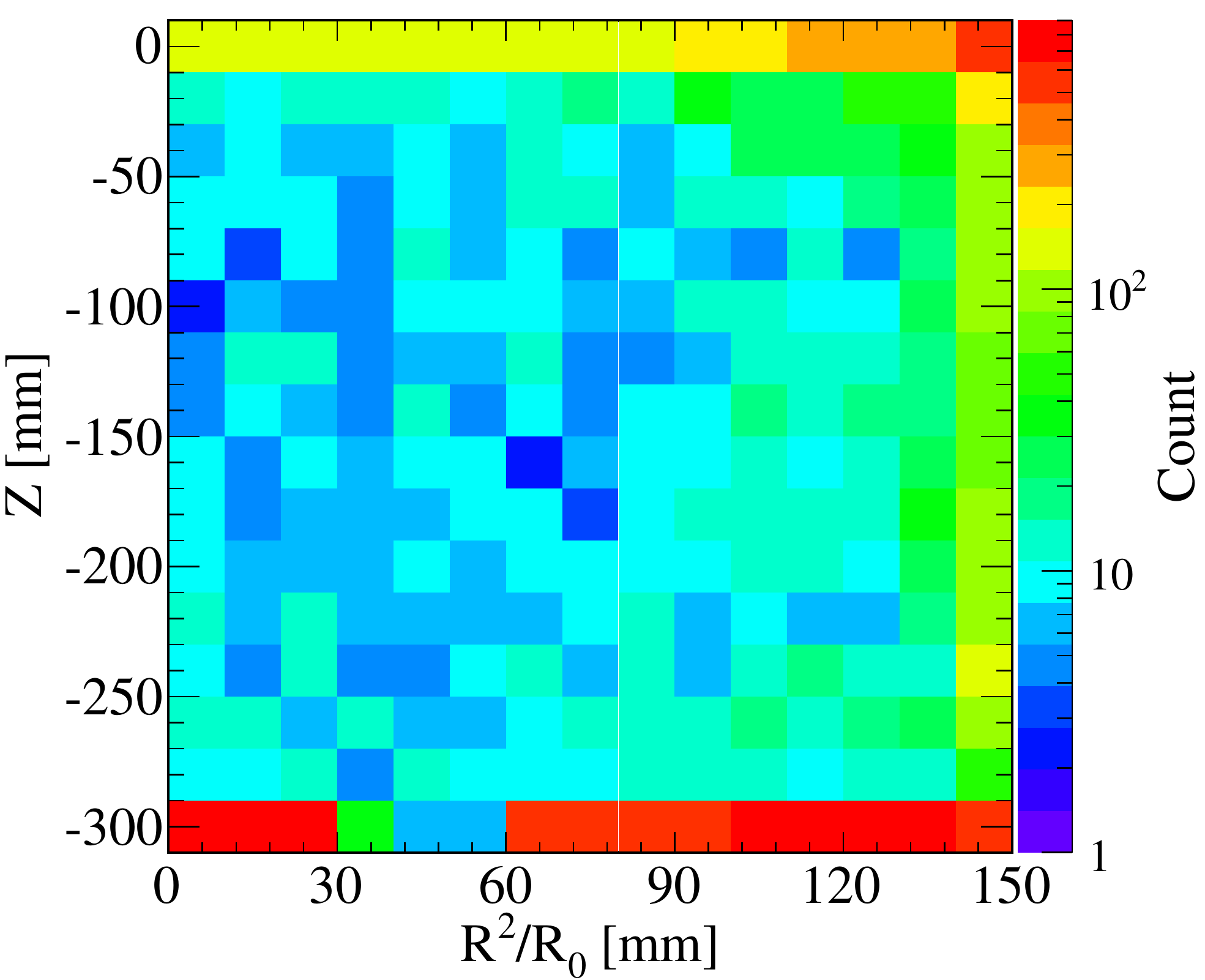}
\includegraphics[width=1.00\columnwidth]{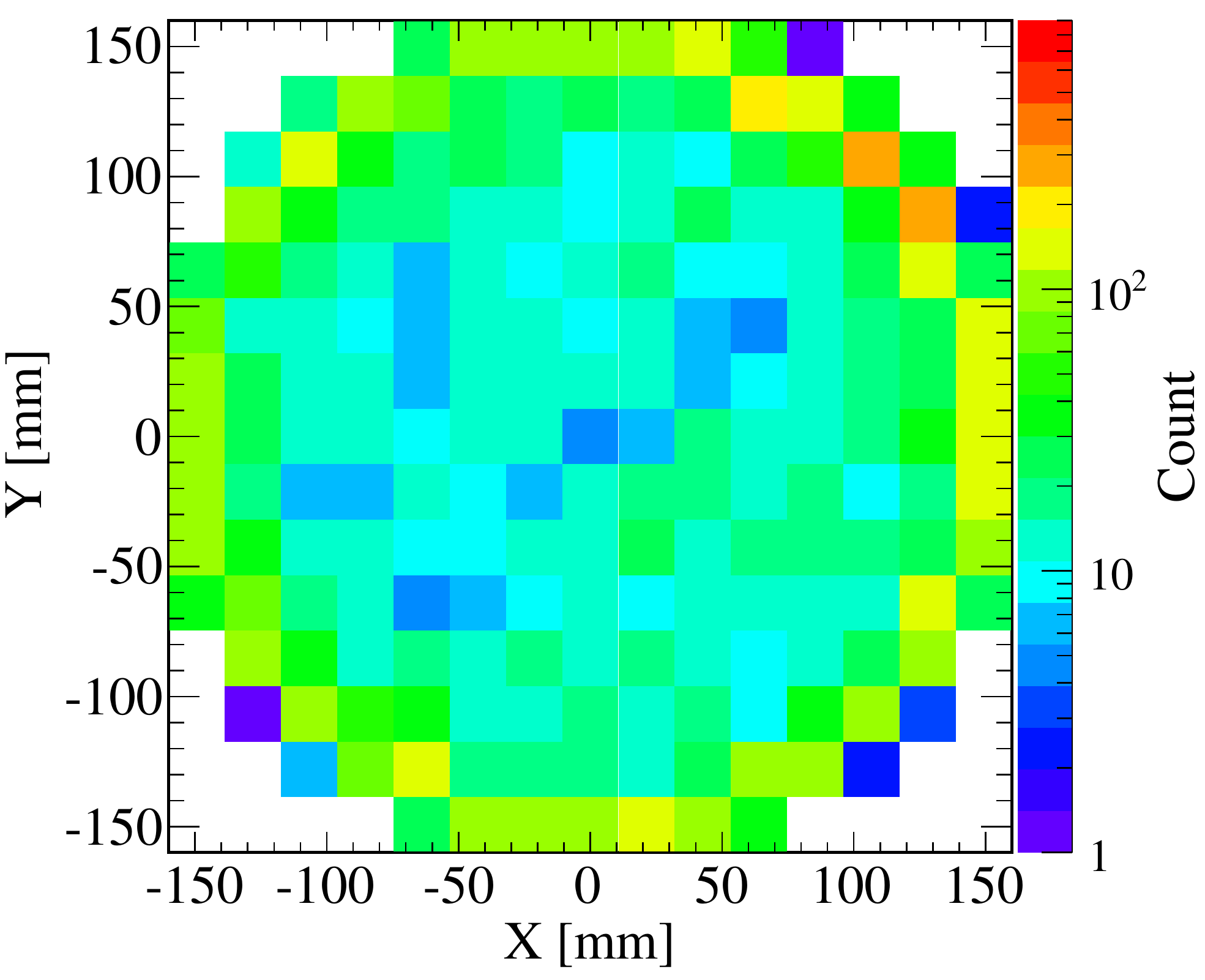}
\caption{The spatial distribution of low-energy events is shown in two different perspectives. Since some clustering near the electrodes is apparent (top), events within $5\1{mm}$ of either electrode have been rejected to display the $XY$-distribution (bottom).}
\label{fig:PbSpatialDist}
\end{figure}

The spatial distribution of these low-energy events is shown in Figure~\ref{fig:PbSpatialDist}. Significant clustering near the electrodes is apparent. The cluster near the anode results from the de-excitation gammas that travel from the veto; whereas low-energy events of ionized $^{212}$Pb populate the cluster at the cathode. For the $XY$ distribution, an additional cut was thus applied to exclude events within $5\1{mm}$ of either the cathode or the anode. As one can see, the low-energy activity is present throughout the TPC. Despite the fact that low-energy $^{212}$Pb events cannot be disentangled from Compton scatters originating in the veto, the distribution of BiPo events in Figure~\ref{fig:BiPoSpatialDist} indicates that the low-energy $^{212}$Pb events reach the center of the active region. The convection pattern observed with the short-lived alpha decays slightly biases this event population toward regions with downward motion. We find 5300 low-energy decays within the central $34\1{kg}$ fiducial region (used in~\cite{Aprile:2012nq}). The aforementioned simulation tells us that 5\% (300) of these events are actually ground-state $^{212}$Pb beta decays, whereas the remainder originates in the veto. Thus, 300 in every $10^6$ events will be useful for low-energy ER calibration of tonne-scale dark matter detectors. This measurement validates the $^{220}$Rn source as a low-energy electronic recoil calibration source for noble element detectors.

Since the full decay chain has a collective decay time less than $12\1{hours}$, the introduced activity decays within a week. Unlike calibration with tritiated methane, this time scale is independent of the purification speed or efficiency, making this source useful for the largest detectors envisioned~\cite{Aalbers:2016jon}.

We observe no activity attributable to either $^{224}$Ra (the  parent of $^{220}$Rn) or $^{222}$Rn after the source is closed and thus place upper limits on the inadvertent introduction of these isotopes. In the case of $^{224}$Ra, the rate of $^{220}$Rn events is compared for periods before and after the source is deployed, resulting in a $^{224}$Ra activity $<1.0\1{\mu Bq/kg}$ at 90\% confidence. The bottom panel of Figure~\ref{fig:TimeEvolution} shows that the rate of $^{222}$Rn remains constant over the period during which the calibration source is deployed, resulting in a $^{222}$Rn activity of $<13\1{\mu Bq/kg}$ at 90\% confidence.

\section{Conclusion}
\label{sec:conclusion}

We have presented a novel calibration method for liquid noble element detectors using a source of dissolved $^{220}$Rn. The $^{220}$Rn decay chain provides several isotopes that allow for a variety of different calibrations, including the response to low-energy beta decays, high-energy alpha lines, and the important $^{222}$Rn background. The activity enters the active volume as soon as the source is opened to the gas purification system. No contamination is observed from long-lived isotopes, and the introduced activity naturally decays within a week after the source is closed. Since this dissipation time is independent of the size of the detector, calibration with $^{220}$Rn is particularly appealing for future large-scale detectors~\cite{Aprile:2015uzo,Akerib:2015cja,Aalbers:2016jon,Amaudruz:2014nsa,Ostrovskiy:2015oja}.

The primary utility of the source is the beta decay of $^{212}$Pb, which can be employed to calibrate a detector's response to low-energy electronic recoil backgrounds in the search for dark matter. The $^{212}$Pb atoms permeate the entire active region, including the center which is beyond the reach of traditional calibrations with external Compton sources.

Furthermore, the high-energy alpha decays of $^{220}$Rn and $^{216}$Po provide the means by which to map atomic motion. We observed a single convection cell in XENON100 at speeds up to $\sim 7\1{mm/s}$ as well as subdominant ion drift in the electric field of the TPC. Such an improved understanding of fluid dynamics within a detector promises to motivate analytic techniques for background mitigation.

Beyond the development of calibration techniques, we have used the beta decay of $^{212}$Bi and the alpha decay of $^{212}$Po to make a high-purity, high-statistics measurement of the half-life of $^{212}$Po: $t_{1/2}=(293.9\pm (1.0)_\text{stat}\pm (0.6)_\text{sys})\1{ns}$.

\section*{Acknowledgements}

We gratefully acknowledge support from the  National Science Foundation, Swiss National Science Foundation, Deutsche Forschungsgemeinschaft, Max  Planck Gesellschaft, Foundation for Fundamental Research on Matter, Weizmann Institute of Science, I-CORE, Initial Training Network Invisibles (Marie Curie Actions, PITNGA-2011-289442), Fundacao para a Ciencia e a Tecnologia, Region des Pays de la Loire, Knut and Alice Wallenberg Foundation, and Istituto Nazionale di Fisica Nucleare. We are grateful to Laboratori Nazionali del Gran Sasso for hosting and supporting the XENON project.

\section*{Appendix: Convection Parameter}
\label{sec:Appendix}

The challenge we face when viewing the convection cell through the lateral surface of the cylindrical TPC is that we see larger subvolumes of the detector closer to the central axis. These unequal volumes could potentially introduce a bias in our measurement of atomic motion. A given subvolume is represented by its projection onto a circle of radius $R_0$ and centered at the origin in the $XY$ plane. The projection is at the position $Y$:

\begin{equation}
dA = 2\sqrt{R_0^2-Y^2}\;dY.
\end{equation}
We aim to convert $A(Y)$ into a function $A(\tilde{Y})$ such that $dA/d\tilde{Y}=\text{constant}$. To this end, we first apply a substitution $Y \equiv -R_0\cos u$, which yields $dA = 2R_0^2\sin^2u~du$. Then, we set

\begin{equation}
\frac{d\tilde{Y}}{du} = \sin^2u = \frac{1}{2}(1-\cos 2u)
\end{equation}
which gives

\begin{equation}
\tilde{Y}(u) = \frac{1}{2}\bigg(u-\frac{1}{2}\sin 2u\bigg)+C.
\end{equation}
To preserve the symmetry around $Y=0$, we require $\tilde{Y}(u(0))=0$ and, thereby, define $C=-\frac{\pi}{4}$. Furthermore, we scale by $4R_0/\pi$,

\begin{equation}
\tilde{Y}(u) = R_0\bigg[\frac{2}{\pi}\bigg(u-\frac{1}{2}\sin 2u\bigg)-1\bigg] \ ,
\end{equation}
so that $\tilde{Y}$ is defined on $[-R_0,R_0]$.

\end{document}